\documentclass[12pt]{iopart}

\usepackage{graphicx}
\begin{document}

\title[G. L. Ma and X.-N. Wang, Initial fluctuations and di- and $\gamma$-hadron correlations]{Initial fluctuations and dihadron and $\gamma$-hadron correlations in high-energy heavy ion collisions}

\author{Guo-Liang Ma$^{1}$ and Xin-Nian Wang$^{2,3}$}
\address{$^{1}$ Shanghai Institute of Applied Physics, Chinese Academy of Sciences, Shanghai 201800, China \\ $^{2}$ Institute of Particle Physics, Central China Normal University, Wuhan 430079, China \\ $^{3}$ Nuclear Science Division MS 70R0319, Lawrence Berkeley National Laboratory, Berkeley, California 94720}
\ead{glma@sinap.ac.cn and xnwang@lbl.gov}

\begin{abstract}
Jets, jet-medium interaction and hydrodynamic evolution of fluctuations in initial parton density all lead to the final anisotropic dihadron azimuthal correlations in high-energy heavy-ion collisions. We remove the harmonic flow background and study the net correlations from different sources with different initial conditions within the AMPT model. We also study $\gamma$-hadron correlations which are only influenced by jet-medium interactions.
\end{abstract}


\vspace{-1.0cm}
\section{Introduction}

Jets, initially produced via hard processes in high-energy heavy-ion collisions, can lose much of their energies during the propagation through the strongly interacting medium that is formed in the collisions~\cite{Wang:1991xy}. Recent experimental data show dihadron azimuthal correlations with double-peak~\cite{Adler:2005ee} and ridge~\cite{Abelev:2009qa} structures, which were thought as medium excitations by or response to jet propagation~\cite{CasalderreySolana:2006qm}. However, initial state is not smooth and calm, but fluctuates. There exist hot spots in the geometrical shapes of the fireball because of the large fluctuations in the initial state. These initial fluctuations lead to many final observables, such as harmonic flow and dihadron correlations. Whether the observed double-peak and ridge structures in dihadron correlations are caused only by harmonic flow (such as triangular flow background)~\cite{Alver:2010gr}or hot spots ~\cite{Takahashi:2009na} is still under debate.

\section{Initial fluctuations and harmonic flow}

\begin{figure}[th]
\center{
\includegraphics[scale=0.25]{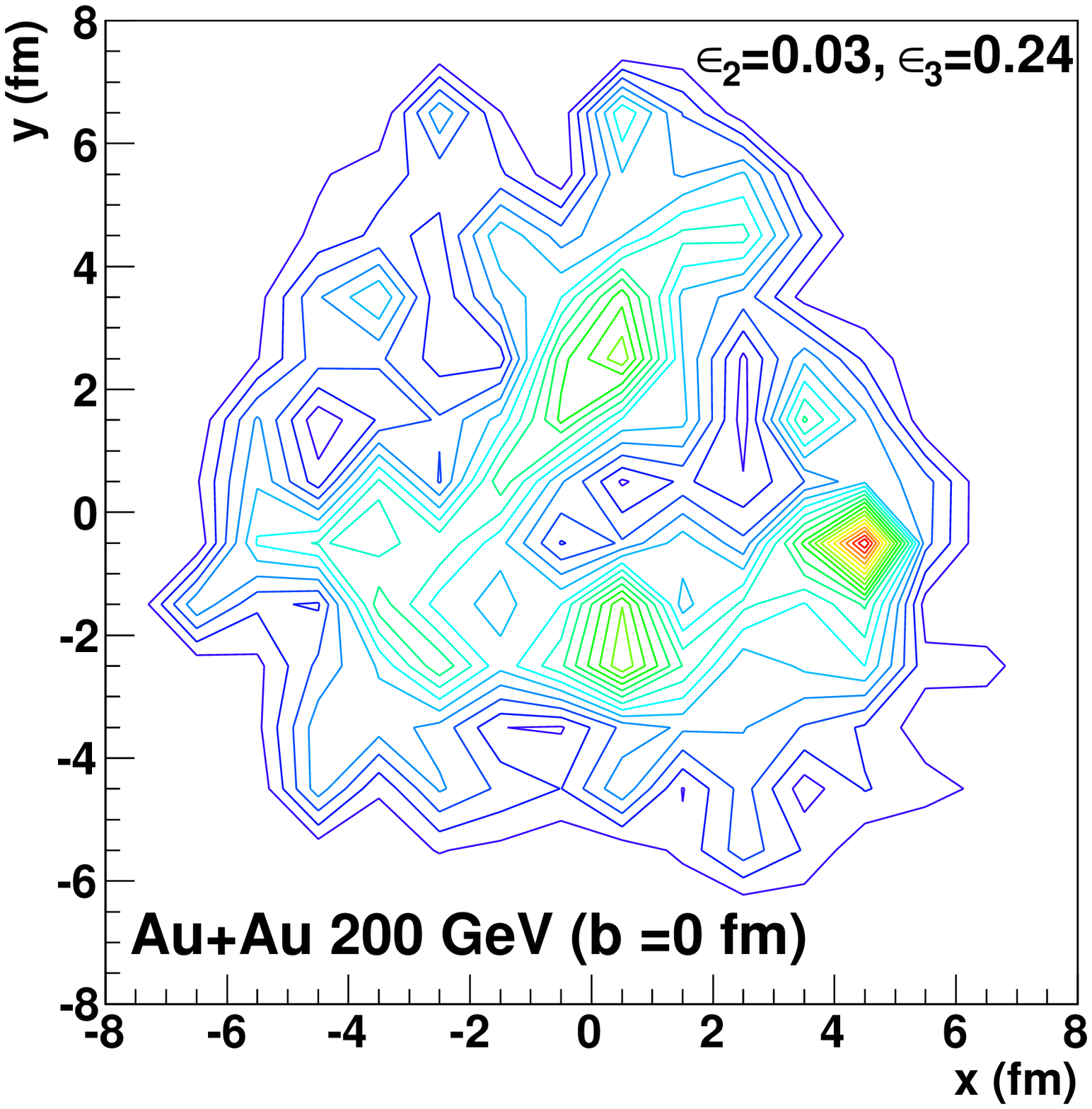}
\includegraphics[scale=0.37]{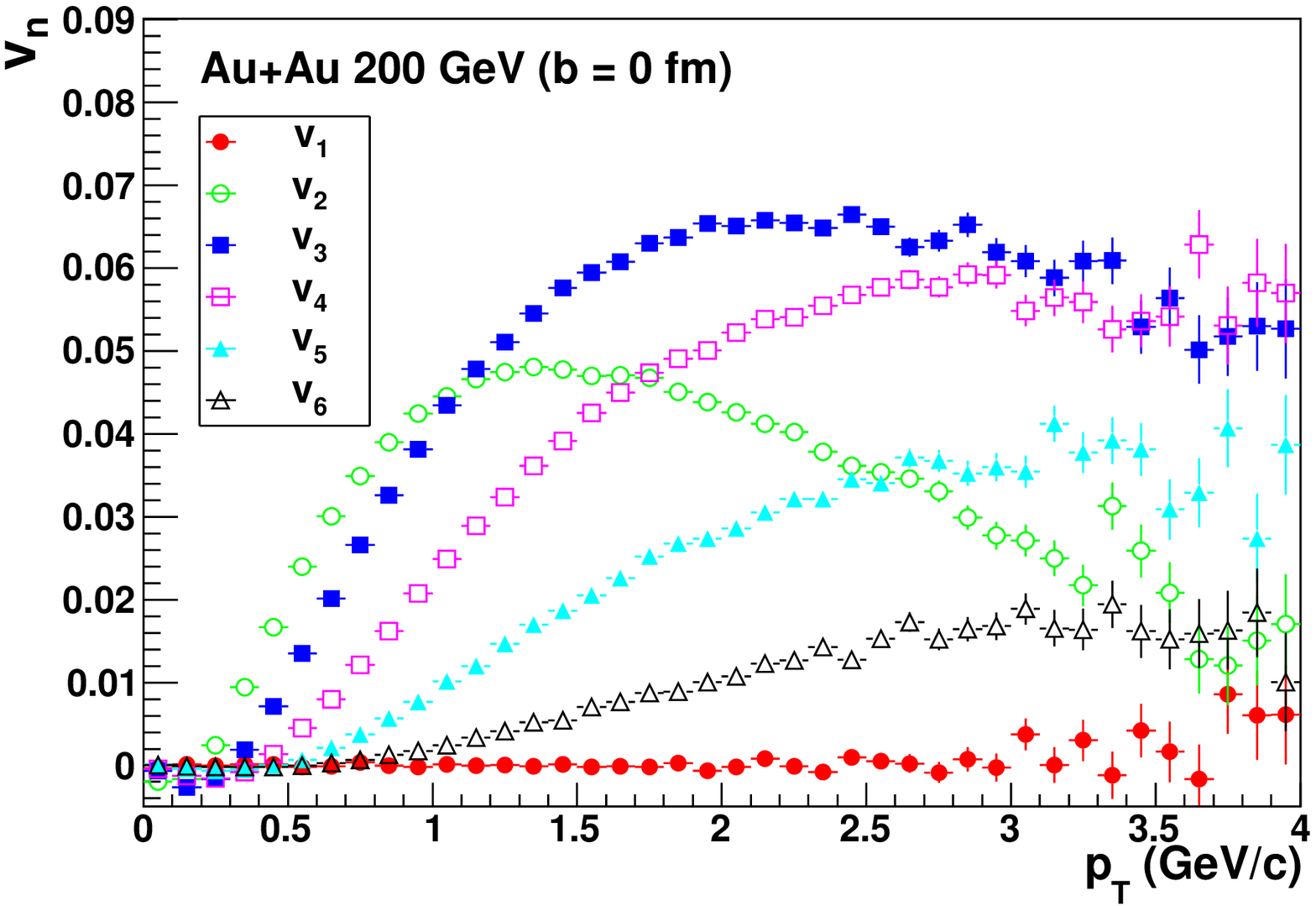}
}
\caption{(Color online) Left panel: Contour plot of initial parton density (in arbitrary unit)  $dN/dxdy$ in transverse plane in a AMPT event; Right panel: Azimuthal anisotropies of hadron spectra $v_{n}(p_{T})$ $(n=1-6)$ from a AMPT model calculation.}
\label{fig1}
\end{figure}

 A Multi-Phase Transport (AMPT) model~\cite{AMPT} is employed to simulate b=0 fm Au+Au 200 GeV collisions (with a partonic interaction cross section of 10 mb) in this study~\cite{Ma:2011ts}. The initial conditions of the AMPT model, including the spatial and momentum distributions of minijet partons and soft string excitations, are obtained from the HIJING model~\cite{HIJING}. Within HIJING, Glauber model for multiple nucleon scattering is used to describe the initial parton production in heavy-ion collisions. Nucleon-nucleon scatterings contain both independent hard parton collisions and coherent soft interactions that are modeled by string formation for each participant nucleon. Strings are then converted into soft partons via string melting scheme in AMPT. Such multiple parton production mechanism leads to fluctuations in local parton number density or hot spots. The left panel in Fig.~\ref{fig1} shows a contour plot of initial parton density in the transverse plane $dN/dxdy$ for a Au+Au event with eccentricity $\epsilon_{2}$=0.03 and triangularity $\epsilon_{3}$=0.24 of the initial transverse parton distribution. It shows the possibility that the initial partons are distributed in a triangular area with many hot spots even for an event with b = 0 fm. The initial geometry asymmetry can be translated into final momentum space by final state interactions, which contribute to all orders of harmonic flow. The right panel in Fig.~\ref{fig1} shows the initial fluctuations can lead to the different orders of harmonic flow ($v_n$, n=1-6),  even for b=0 fm Au+Au 200 GeV collisions.

\section{Dihadron correlations}

\begin{figure}[th]
\includegraphics[scale=0.34]{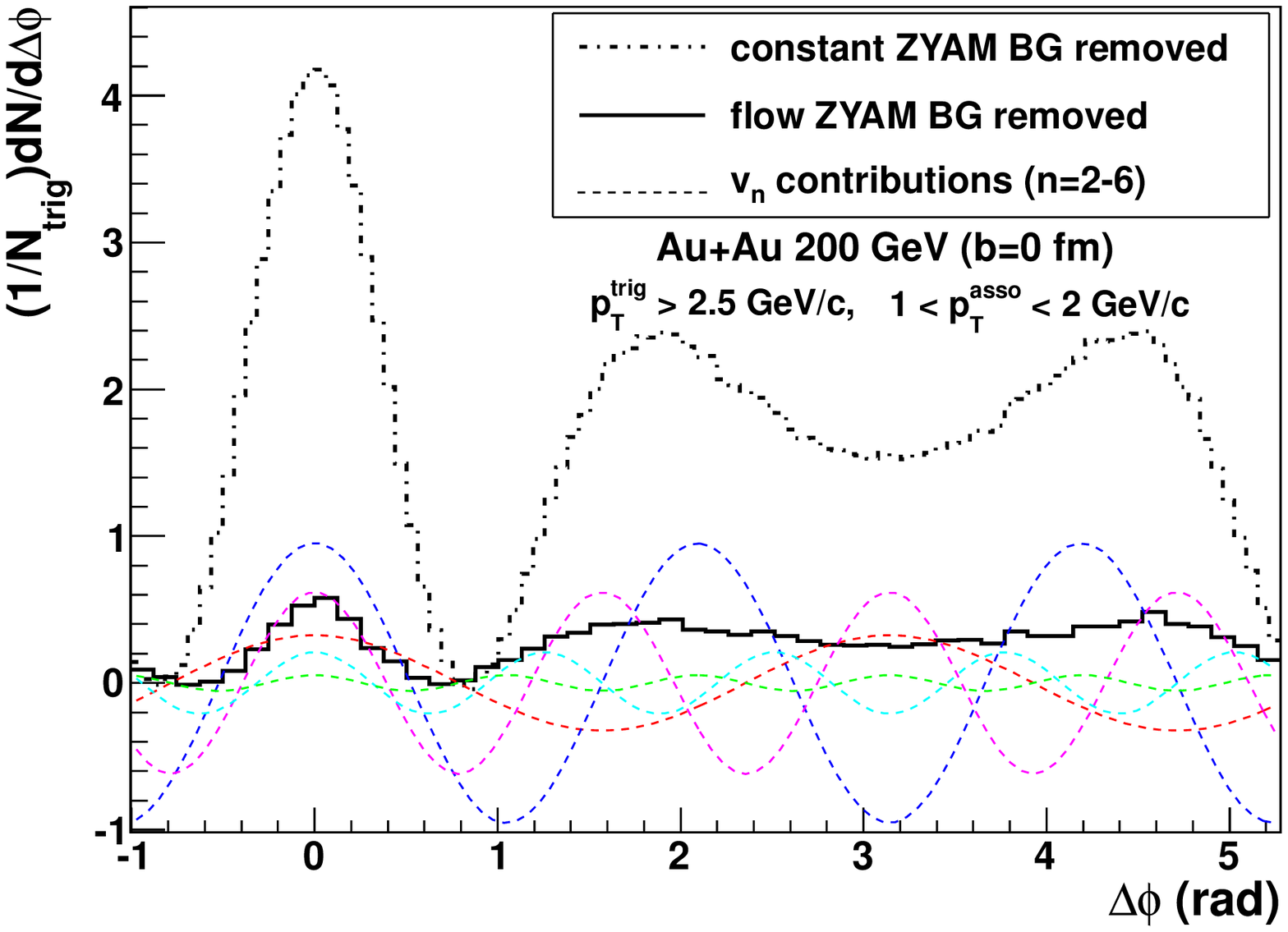}
\includegraphics[scale=0.35]{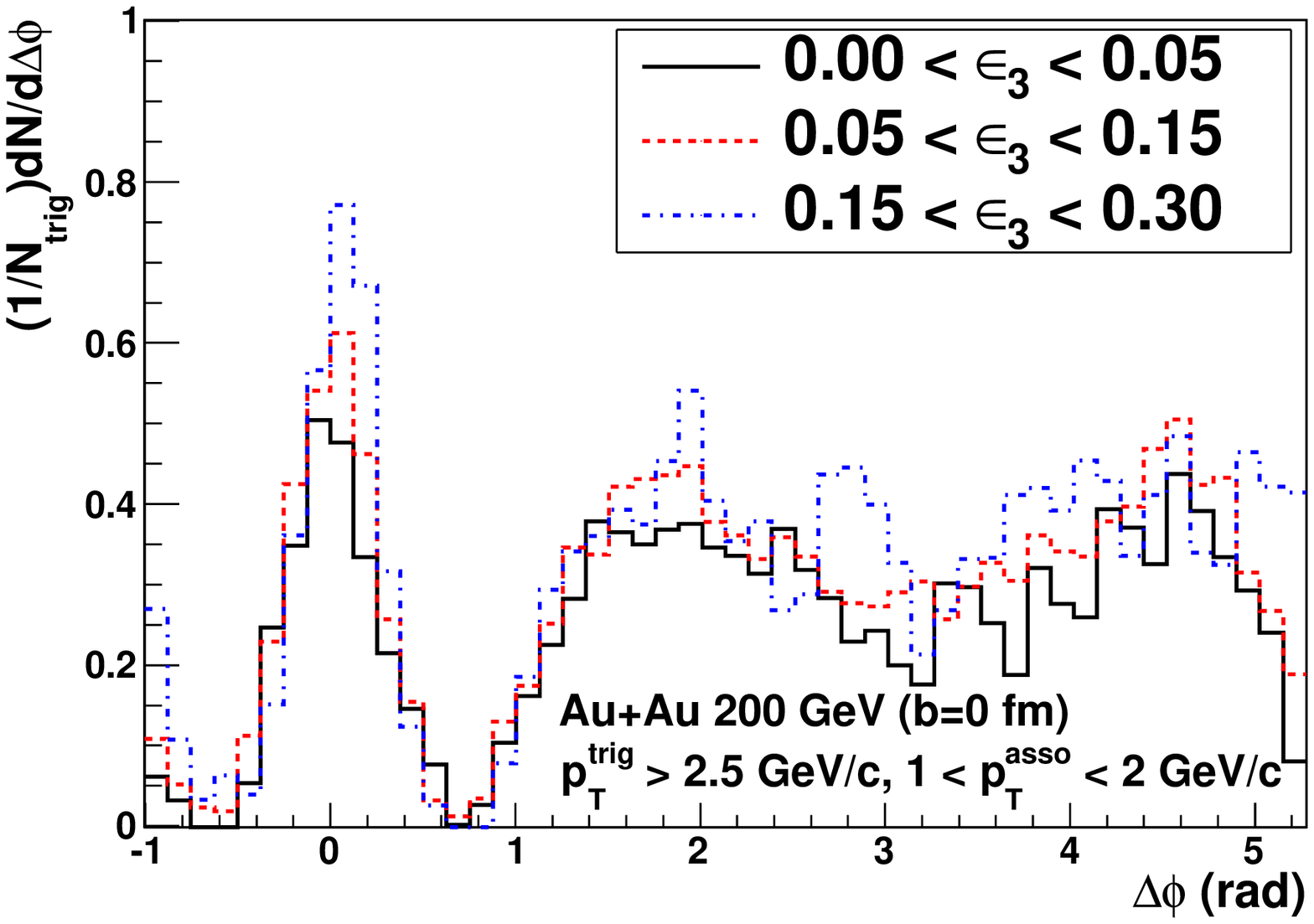}
\caption{(Color online) Left panel: AMPT results on dihadron correlations before (dot-dashed) and after (solid) subtraction of the contribution from harmonic flow $v_{n} (n=2-6)$; Right panel: AMPT results on dihadron correlations after subtraction of harmonic flow with different values of the geometric triangularity $\epsilon_{3}$. }
\label{fig2}
\end{figure}

To find contribution to dihadron correlations from other mechanisms such as jet-medium interactions, it is very important to remove the background contribution from all orders of harmonic flow. We subtract the flow contribution,
$f(\Delta \phi ) = B\left(1 + \sum\limits_{n = 1}^\infty {2\langle v_n^{\rm trig} v_n^{\rm asso}\rangle \cos n\Delta \phi} \right)$
from dihadron correlations, where B is a normalization factor determined by the ZYAM scheme, $v_n^{trig}$ and $v_n^{asso}$ are harmonic flow coefficients for trigger and associated hadrons. The left panel in Fig.~\ref{fig2} shows dihadron correlations before (dot-dashed) and after (solid) the removal of
contributions from harmonic flow [$v_n$, $n=2$-6 (dashed)]. As shown in the right panel in Fig.~\ref{fig2}, the dihadron correlations with different initial geometric triangularity $\epsilon_{3}$ become identical after the harmonic flow background subtraction, which indicates that we have removed the complete harmonic flow background.

\begin{figure}[th]
\center{
\includegraphics[scale=0.35]{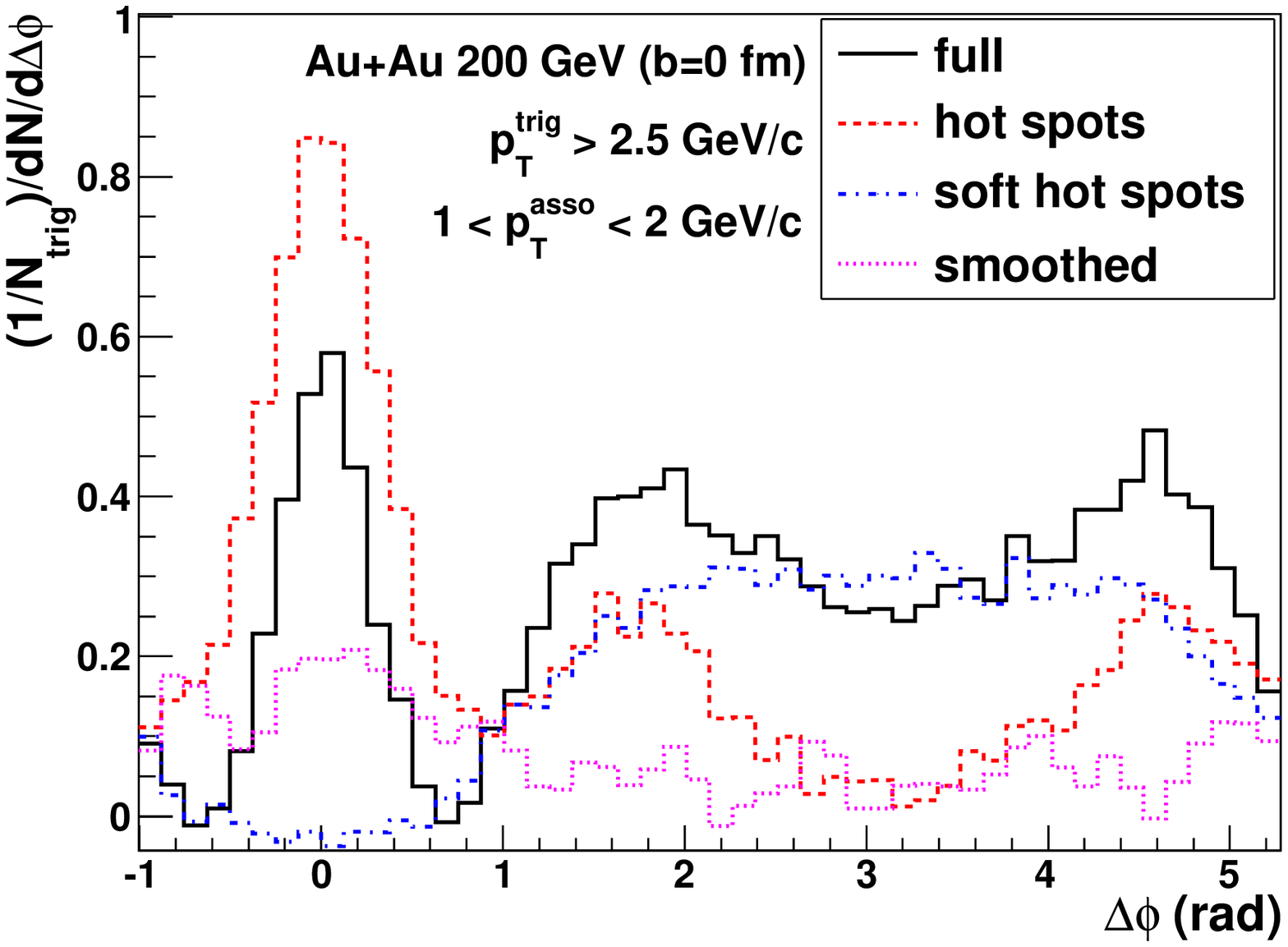}
\includegraphics[scale=0.35]{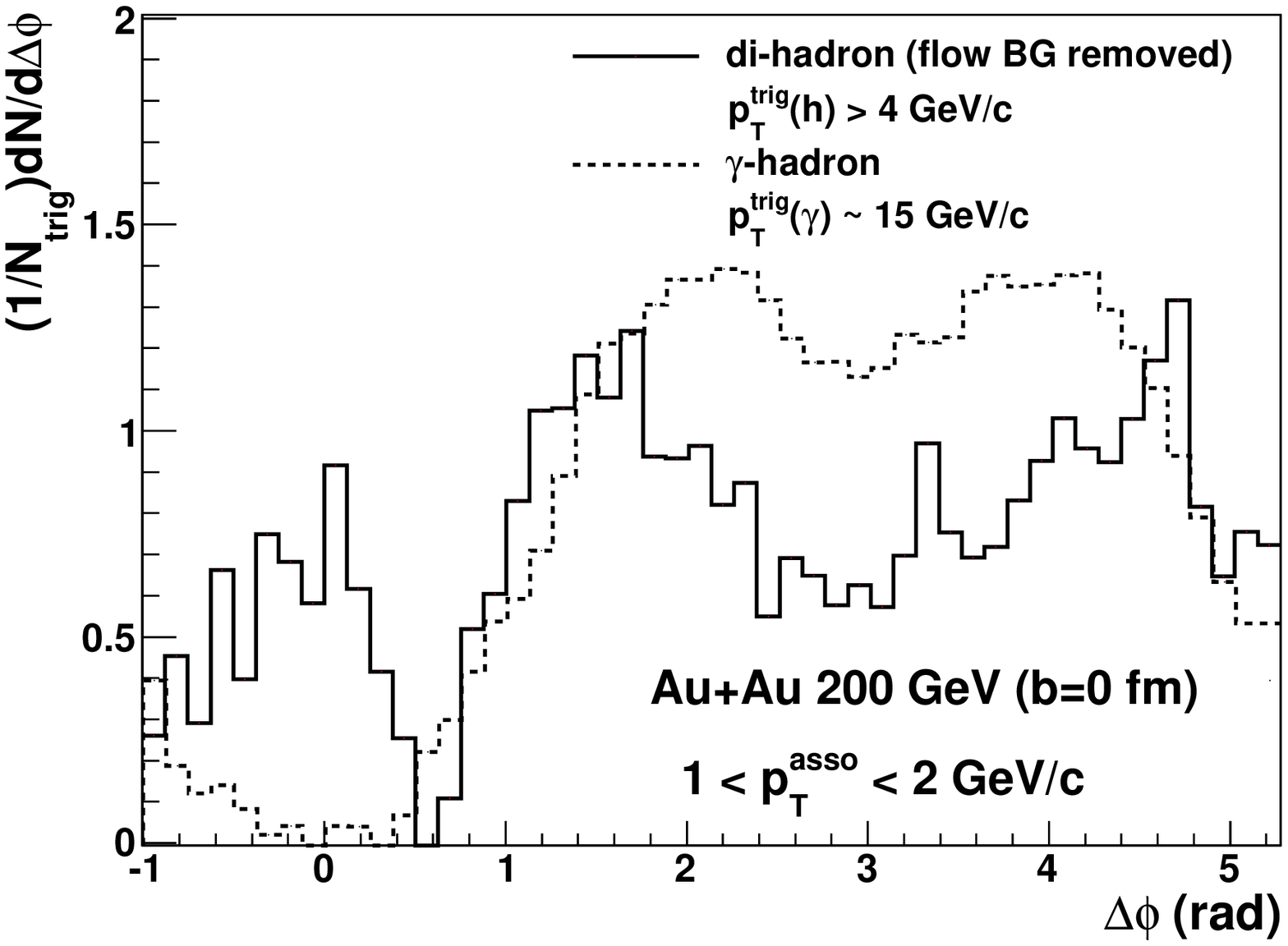}
}
\caption{Left panel: (Color online) Dihadron correlations (with harmonic flow subtracted) from AMPT calculations with different initial conditions; Right panel: Dihadron correlation (solid) compared with $\gamma$-hadron correlation (dashed) from AMPT model calculations.}
\label{fig3}
\end{figure}

Four different initial conditions are adopted to study separately the effects of jets and hot spots on dihadron correlations within AMPT model. We first randomize the azimuthal angle of each jet shower parton in the initial condition from HIJING simulations. This effectively switches off the initial back-to-back correlations of dijets. The dihadron correlation (dashed) denoted as ``hot spots'' in the left panel of Fig.~\ref{fig3} still exhibits a double-peak on the away-side that comes only from hot spots. It has roughly the 
same opening angle $\Delta\phi\sim 1$ (rad) as in the ``full'' simulation (solid).  However, the magnitude
of double peaks is reduced, which can be attributed to medium modified dijets and jet-induced medium excitations. When jet production is turned off in the HIJING initial condition,  fluctuations in soft partons from strings can still form  what we denote as ``soft hot spots'' that lead to a back-to-back dihadron correlation (dot-dashed) with a broadening peak. Without jets in AMPT, one can further randomize the polar angle of  transverse coordinates of soft partons and therefore eliminate the ``soft hot spots''. The dihadron correlation from such ``smoothed'' initial condition becomes almost flat (dotted). We here emphasize that the above results include both short-range and long-range dihadron correlations ($\Delta\eta<$ 2). We observed that hot spots are elongated in longitudinal direction as tubes, which can finally contribute to the formation of ridge structure up to $\Delta\eta\sim$ 2.5. 

\section{$\gamma$-hadron correlations}
 
 $\gamma$-hadron correlations are golden probes to study jet-medium interactions. Direct photons are produced isotropically, have neither strong interactions with medium nor harmonic flow, therefore the backgrounds for $\gamma$-hadron correlations are flat. Any structure in $\gamma$-hadron correlations with large $p_{T}^{\gamma}$ should be only due to jet-medium interactions. The right panel in Fig.~\ref{fig3} shows $\gamma$-hadron correlation is comparable with dihadron correlation in magnitude but with a less pronounced double-peak, which can be attributed to additional dihadron correlations from hot spots, geometric bias toward surface, and tangential emissions that enhance deflections of jet showers and jet-induced medium excitations by radial flow~\cite{Li:2011ts}.

\section{Conclusions}

The fluctuations of initial geometry lead to different orders of harmonic flow, which can significantly affects dihadron azimuthal correlations. After removing harmonic flow background, the net dihadron correlations reflect hot spots and jet-medium interactions. $\gamma$-hadron correlations are proposed as golden probes because they only come from jet-medium interactions.


\ack{
This work is supported by the NSFC of China under Projects Nos. 10705044, 10875159, 10905085, 10975059, 11035009, the Knowledge Innovation Project of Chinese Academy of Sciences under Grant No. KJCX2-EW-N01 and by the U.S. DOE under Contract No. DE-AC02-05CH11231 and within the framework of the JET Collaboration.
}

\end{document}